\documentclass{iau} 
\usepackage{graphicx}
\usepackage{natbib}
\usepackage{subfigure}
\usepackage[version=4]{mhchem}

\newcommand{\farcs}{\mbox{\ensuremath{.\!\!^{\prime\prime}}}}%  % fractional arcsecond symbol: 0.''0

\title[COMs in disks with ALMA]
{Methanol formation in TW~Hya and future prospects for detecting larger complex molecules in disks with ALMA}

\author[Walsh, C., et al.]
{Catherine Walsh$^{1,2}$, Shreyas Vissapragada$^{2,3}$, \and Harry McGee$^1$}

\affiliation{$^1$School of Physics and Astronomy \\ University of Leeds, Leeds LS2 9JT, UK \\ 
email: {\tt c.walsh1@leeds.ac.uk} \\[\affilskip]
$^2$Leiden Observatory \\ Leiden University, P.O.~Box 9513, Leiden 2300 RA, Netherlands \\ [\affilskip]
$^3$Columbia Astrophysics Laboratory \\ Columbia University, New York, NY 10027, USA}

\pubyear{2017}
\volume{332}
\jname{Astrochemistry VII - Through the Cosmos from Galaxies to Planets}
\editors{M.R.~Cunningham, T.J.~Millar \& Y.~Aikawa, eds.}

\begin{document}

\maketitle

%% add a maximum of 10 keywords, to be taken form the file <Keywords.txt>
\begin{abstract}
Gas-phase methanol was recently detected in a protoplanetary disk for the first time with ALMA.  
The peak abundance and distribution of methanol observed in TW Hya differed from that 
predicted by chemical models.
Here, the chemistry of methanol gas and ice is calculated using a physical model 
tailored for TW Hya with the 
aim to contrast the results with the recent detection in this source. 
New pathways for the formation of larger complex molecules (e.g., ethylene glycol) are included 
in an updated chemical model, as well as the fragmentation of methanol ice upon photodesorption.  
It is found that including fragmentation upon photodesorption 
improves the agreement between the peak abundance reached in 
the chemical models with that observed in TW Hya ($\sim 10^{-11}$ with respect to \ce{H2}); 
however, the model predicts that the peak in emission resides a factor of $2-3$ farther 
out in the disk than the ALMA images.  
Reasons for the persistent differences in the gas-phase methanol distribution 
between models and the observations of TW Hya are discussed. 
These include the location of the ice reservoir which may coincide with 
the compact mm-dust disk ($\lesssim 60$~au)   
and sources of gas-phase methanol which have not yet been considered in models.
The possibility of detecting larger molecules with ALMA is also explored.  
Calculations of the rotational spectra of complex molecules other than methanol using a parametric 
model constrained by the TW Hya observations suggest that the detection of individual emission lines 
of complex molecules with ALMA remains challenging.  However, the signal-to-noise ratio can be enhanced 
via stacking of multiple transitions which have similar upper energy levels.

\keywords{astrochemistry, planetary systems: protoplanetary disks, molecular processes, stars: individual (TW Hya)}
\end{abstract}

% if your document starts with a section, remove some space above using this command.
\firstsection 

\section{Introduction}              
The detection of complex organic molecules (COMs) in protoplanetary disks has been considered 
a key science goal for the Atacama Large Millimeter/submillimeter Array (ALMA).  
In astrochemistry, a COM is loosely defined as a carbon-containing molecule containing 
six or more atoms \citep[see, e.g.,][]{herbst09}.  
Many families of COMs (e.g., amines, aldehydes, and carboxylic acids) are thought to be 
stepping stones towards prebiotic molecules such as amino acids and simple sugars.  
COMs typically possess a complex rotational spectrum; hence, 
emission from COMs is weak, especially from small astrophysical objects such as 
protoplanetary disks ($\sim 1\farcs$).  
An additional challenge for protoplanetary disks in particular is that most of the disk 
material is cold ($\lesssim 100$~K) and COMs are expected to reside on and within the 
water-dominated ice mantles on dust grains.  
However, the detection of cold gas-phase water and ammonia in the nearby disk around TW~Hya 
has confirmed that non-thermal desorption of ice species into the gas phase does occur in disks. 
This helps to maintain a low, yet detectable, amount in the gas \citep{hogerheijde11,salinas16}.  

The superior sensitivity and spatial resolution of ALMA has allowed, for the first time, 
the possibility to see relatively complex molecules in nearby protoplanetary disks \citep[e.g.,][]{oberg15}.  
Such detections are vital for probing the link between the chemical complexity 
observed towards the warm and dense gas in the immediate environment of 
forming stars \citep[e.g.,][]{herbst09,caselli12} and that seen in comets in the 
Solar System \citep[e.g.,][]{mumma11,leroy15}.  
It remains unclear to what extent chemistry during disk formation 
and evolution shapes the chemical complexity of icy planetesimals.  
Recent theoretical models of complex chemistry during disk formation 
and over the disk lifetime have suggested that (i) chemical complexity 
in ices is enhanced en route into the disk \citep{drozdovskaya14,drozdovskaya16,yoneda16}, 
and (ii) gas-phase COMs should be present in sufficiently detectable quantities 
in nearby protoplanetary disks \citep[e.g.,][]{furuya14,walsh14,parfenov16}.  
Methanol is an abundant interstellar ice \citep[see, e.g.,][]{boogert15} 
which has been shown in the laboratory to be an important 
feedstock for larger and more complex molecules 
\citep[e.g.,][]{oberg09,modica10,chen13}; 
hence, gas-phase methanol 
is a key target for dedicated observational studies in nearby 
protoplanetary disks.  

\section{Detection of gas-phase methanol in TW Hya}

Four rotational transitions of A-type methanol were targeted  
during ALMA Cycle 2 observations of TW Hya \citep{walsh16}.  
This spectral set-up, within which four strong lines could be 
simultaneously observed, was motivated by simulations of 
both the chemical structure and the emergent spectrum 
across the full ALMA frequency range \citep{walsh14}.  
This work predicted that methanol emission from the 
cold, outer disk would dominate, arising from photodesorption 
of the ice reservoir.  
Hence, the targeted lines possessed relatively low-lying upper level 
energies ranging from $\approx 20$~K to $\approx 100$~K.

Data reduction and imaging was conducted as described in 
\citet{walsh16}.  
The individual lines were not detected in the synthesised images; 
however, a detection was made possible through stacking 
of the three transitions at 
304.208~GHz ($E_\mathrm{up} = 21.6$~K), 
305.473~GHz ($E_\mathrm{up} = 28.6$~K), and 
307.166~GHz ($E_\mathrm{up} = 38.0$~K).  
The gas-phase methanol was detected with a signal-to-noise ratio 
of 5.5 in the stacked channel maps, reaching a sensitivity of 
2~mJy~beam$^{-1}$ in a 0.15~km~s$^{-1}$ channel.  
A more recent analysis of this dataset using a matched filter to 
search for emission in the $uv$ domain showed 
that the three lines listed above were robustly detected with 
signal-to-noise levels of 4.4, 6.2, and 3.4, respectively 
\citep{loomis17}.

\begin{figure}[]
\begin{center}
\includegraphics[width=0.5\textwidth]{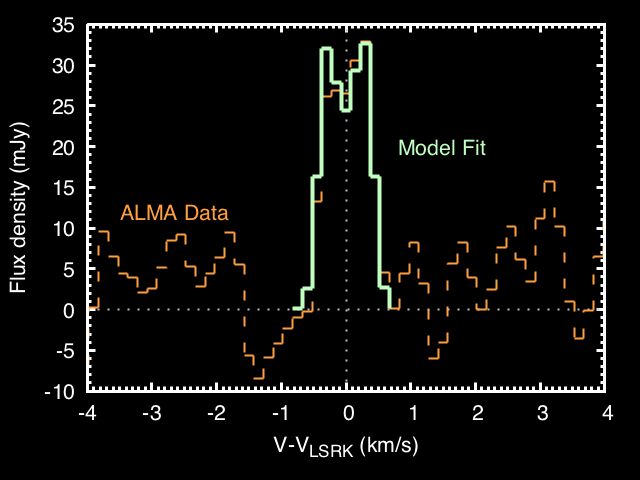} 
\caption{Methanol line profile from TW Hya extracted from the stacked 
channel maps within the $3\sigma$ contour of the 317~GHz continuum 
emission (orange dashed line) compared with the model fit line profile (green solid line). 
Figure adapted from \citet{walsh16}.}
\label{fig1}
\end{center}
\end{figure}

The stacked emission was then simulated using a physical model 
of TW Hya constrained by a wealth of observations \citep{kama16} 
and using a parametric model for the gas-phase methanol distribution 
and abundance.
This suggested that the gas-phase methanol arises in ring, peaking at 
$\approx 30$~au, and with an abundance ranging from 
$3 \times 10^{-12} - 4 \times 10^{-11}$ with respect to \ce{H2}.  
The derived emission morphology and abundance differed from 
that predicted by previous models.  
First, the methanol emission in TW Hya appeared more compact 
than the predictions, 
and second, the peak abundance was $2-3$ orders of magnitude 
lower than expected \citep{walsh14}.  
Figure~\ref{fig1} shows the line profile in orange dashed 
lines with the best-fit model overlaid in green.
 
\section{An updated chemical model for methanol gas and ice}

The apparent discrepancy between predictions and the observations 
of methanol in TW Hya may be reconciled by modelling the complex chemistry 
using a disk structure specific to TW Hya.  
TW Hya is a colder and more settled disk that that used for the 
previous predictions \citep[from][]{nomura07}.  
In addition, since the models presented in \citet{walsh14}, there have been 
numerous laboratory investigations into methanol ice chemistry. 
Studies into methanol ice photodesorption showed that methanol 
does not desorb intact at low temperatures \citep{bertin16,cruzdiaz16}.  
The upper limit for the methanol photodesorption was found to be 
$\lesssim 10^{-5}$ molecules~photon$^{-1}$ for methanol mixed with CO. 
This alone could help to reconcile the low methanol abundances 
seen in TW Hya because previous estimates of intact methanol photodesorption suggested 
a yield of $\sim 10^{-3}$ molecules photon$^{-1}$ \citep{oberg09}.  

To investigate the abundance and distribution of methanol gas and ice 
an updated chemical model was developed and the chemistry was computed 
using an observationally constrained physical model of the TW Hya 
protoplanetary disk \citep[from][]{kama16}. 
The chemical model presented in \citet{walsh15} was updated 
with the fragmentation pathways for methanol ice photodesorption \citep{bertin16}.   
The current model also includes an extended surface chemistry network for methanol 
and its related compounds from \citet{chuang16} including methyl formate 
(\ce{HCOOCH3}), glycolaldehyde (\ce{HOCH2CHO}), glyoxal (\ce{HC(O)CHO}), 
and ethylene glycol (\ce{(CH2OH)2}).  

\begin{figure}[]
\begin{center}
\includegraphics[width=\textwidth]{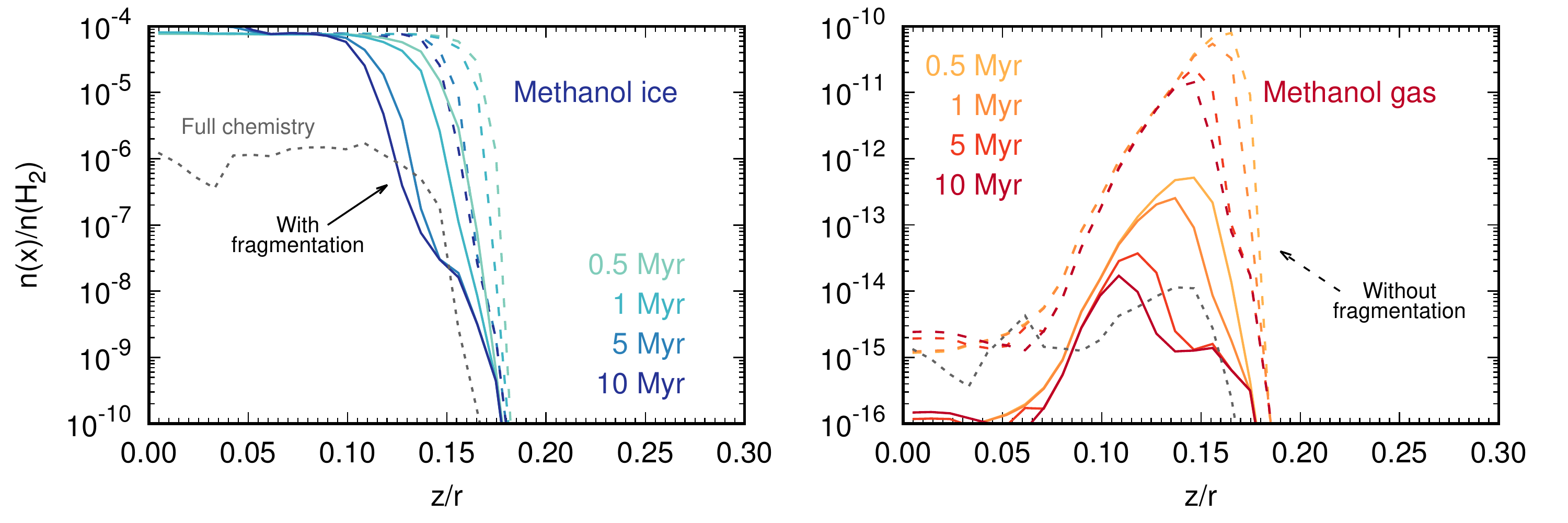} 
\caption{Fractional abundance of methanol ice (left) and gas (right) 
with respect to \ce{H2} as a function of disk vertical height 
($z/r$) and at a radius of 30~au. The results 
with (solid) and without (dashed) fragmentation 
upon photodesorption are shown.  
The y-axis scales differ due to the low abundance of the gas 
relative to the ice.  
Results at four time steps are plotted (0.5, 1.0, 5.0, and 10~Myr). 
The dotted gray lines show the results at 1 Myr using a more comprehensive chemical 
model (see text for details).}
\label{fig2}
\end{center}
\end{figure}

Figure~\ref{fig2} shows the fractional abundances of methanol ice (left) 
and gas (right) as a function of disk height, $z/r$, at a radius 
of 30~au and at four different time steps 
(0.5, 1.0, 5.0, and 10~Myr) for a model in which photodesorption only 
is included as a non-thermal desorption mechanism.  
More details on these calculations are presented in an 
upcoming publication by \citet{ligterink17}.  
When fragmentation upon photodesorption is included, the gas-phase abundance drops by 
$2-3$ orders of magnitude in the molecular layer present at 
the methanol snow surface ($z/r \approx 1.5$).  
The snow surface marks the boundary in the disk atmosphere at which a molecule 
transitions from the ice phase to the gas phase.  
Further, the location of the methanol snow surface 
becomes deeper in time: this is because the methanol ice now needs 
to reform from its fragments rather than simply refreeze from the 
gas phase once desorbed.  

\section{The chemical origin of gas-phase methanol in TW Hya}

\begin{figure}[]
\subfigure{\includegraphics[width=0.5\textwidth]{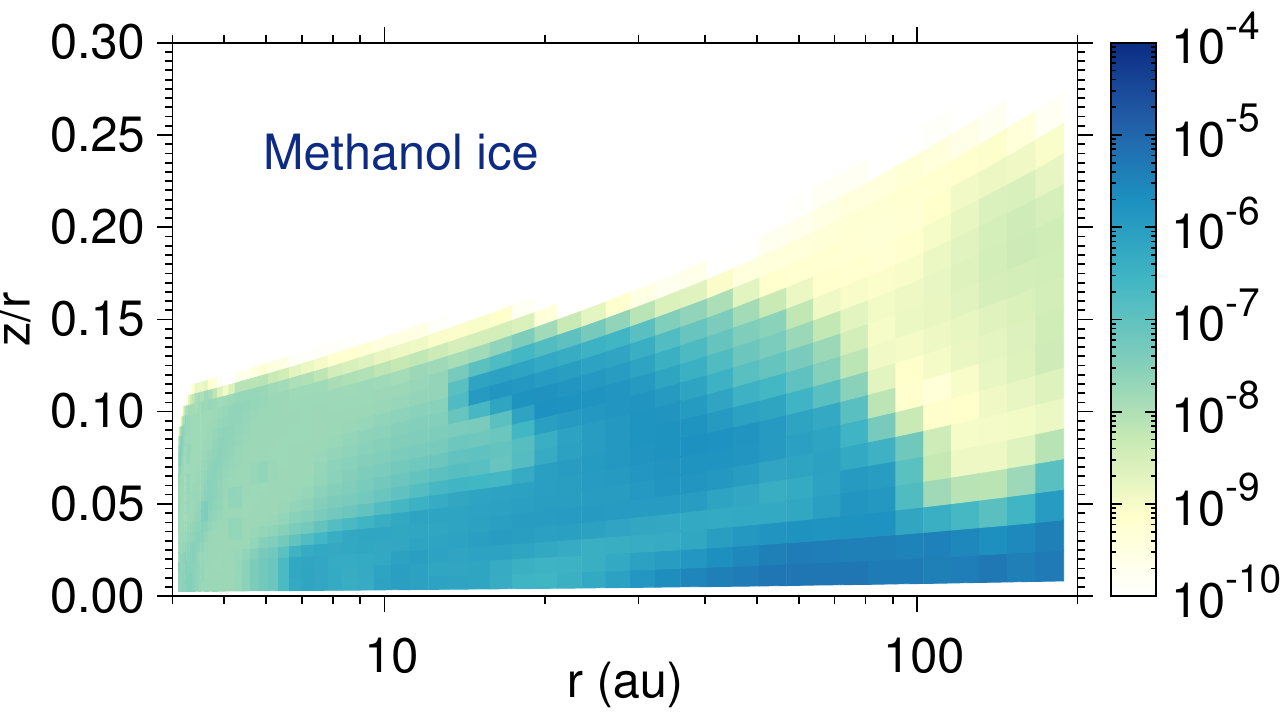}}
\subfigure{\includegraphics[width=0.5\textwidth]{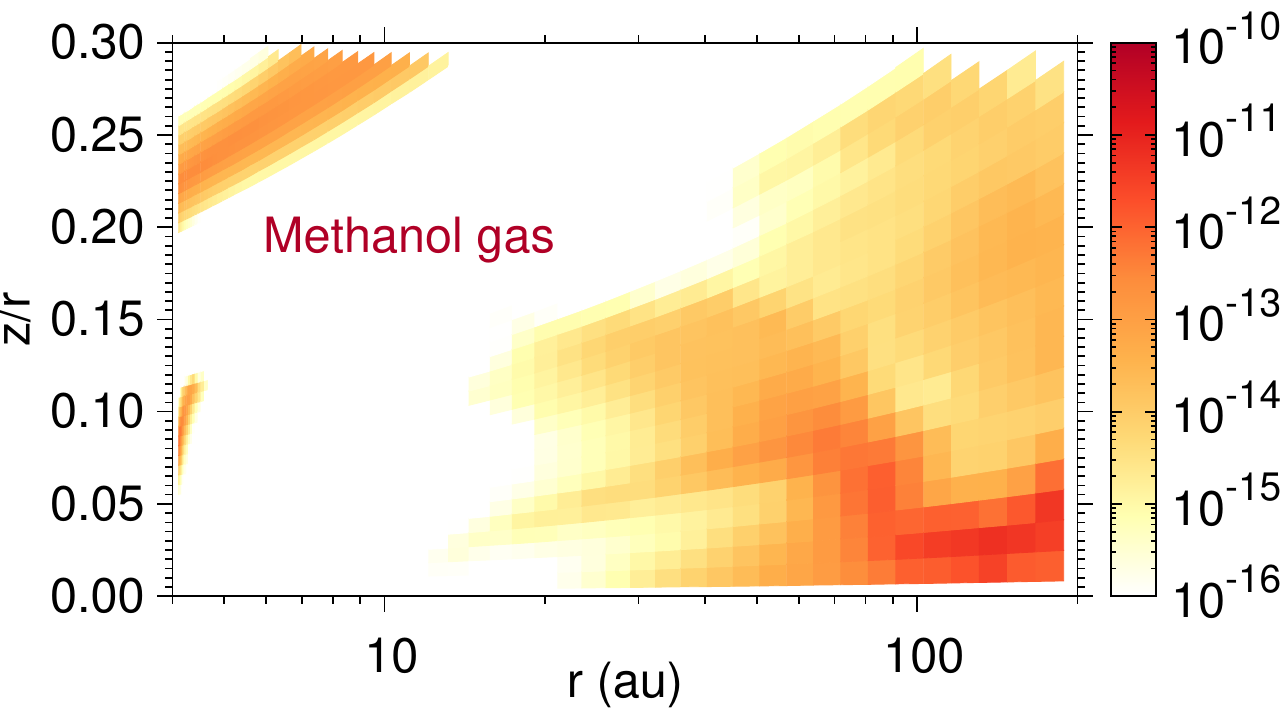}}
\caption{Fractional abundance of methanol ice (left) and gas (right) with respect to 
\ce{H2} as a function of disk radius and height (scaled by the radius) and at a time of 
1~Myr.}
\label{fig3}
\end{figure}

In Figure~\ref{fig3} the two-dimensional fractional abundances of methanol ice 
(left) and gas (right) are shown as a function of disk 
radius, $r$, and height scaled by the radius, $z/r$.  
These have been computed using the updated chemical model as described previously.  
Note that the colour-bar scales differ due to the difference 
in peak abundance between the gas and the ice.  
This model is more complex than that used for the results presented in 
Figure~\ref{fig2} in that it also includes all sources of non-thermal 
desorption (photodesorption and reactive desorption) 
as well as chemistry induced by bulk ice photodissociation 
\citep[see][for full details]{walsh14}.  
The fractional abundances are shown at a single time step (1~Myr).  

As expected, the methanol ice is confined to the midplane throughout the 
disk, lying below $z/r \approx 0.1-0.15$ within 100~au and 
below $z/r\approx 0.05$ beyond 100~au. 
This is due to the increasing penetration of stellar and interstellar UV 
radiation in the outer disk.
The gas-phase methanol has a different distribution to that in previous models 
that predicted that it resides in the outer disk atmosphere 
\citep[$z/r \gtrsim 0.2$,][]{semenov11,furuya14,walsh14}. 
The peak abundance of $\sim 10^{-11}$ with respect to \ce{H2} is reached in the 
outer disk beyond $\approx 100$~au with an abundance of $\gtrsim 10^{-12}$ 
reached beyond $\approx 80$~au.  
The bulk of the gas-phase methanol also lies close to the 
midplane ($z/r \lesssim 0.15$) and mirrors the distribution of 
methanol ice by becoming more confined to the midplane ($z/r \lesssim 0.05$) 
beyond $\approx 100$~au.  
The peak abundance reached is of the same order of magnitude as 
that derived from the observations suggesting that the inclusion of 
fragmentation of methanol ice upon photodesorption does help to 
reduce the peak gas-phase abundance in the outer disk.  
The distribution of methanol ice and gas in this model are very similar to that 
presented in recent work by \citet{parfenov17} for the case where they assume 
that the barrier for diffusion (across the grain/ice surface) is 0.4 times that for 
desorption.  Here we use a value of 0.3.

Figure~\ref{fig2} shows how the more complex chemical model results (dotted lines) 
compare with the more simple models which neglect 
bulk ice photodissociation and reactive desorption (solid and dashed lines).  
The ice abundance 
is lower in the midplane ($\sim 10^{-6}$ with respect to \ce{H2}) 
demonstrating the impact of ice photodissociation on the processing of 
feedstock ice species such as methanol.  
The gas-phase distribution demonstrates that (i) the combination of reactive desorption and 
photodesorption does not help to boost the peak gas-phase abundance of methanol in the 
molecular layer ($z/r \approx 0.1-0.15$) and (ii) reactive desorption does help boost the abundance in 
the midplane ($z/r\lesssim0.1$) to higher than that from models which include photodesorption 
(with fragmentation) only. 
The overall lower peak abundance reached in the more complex model is related 
to the depletion of methanol ice over time.  

\begin{figure}[]
\begin{center}
\includegraphics[width=0.5\textwidth]{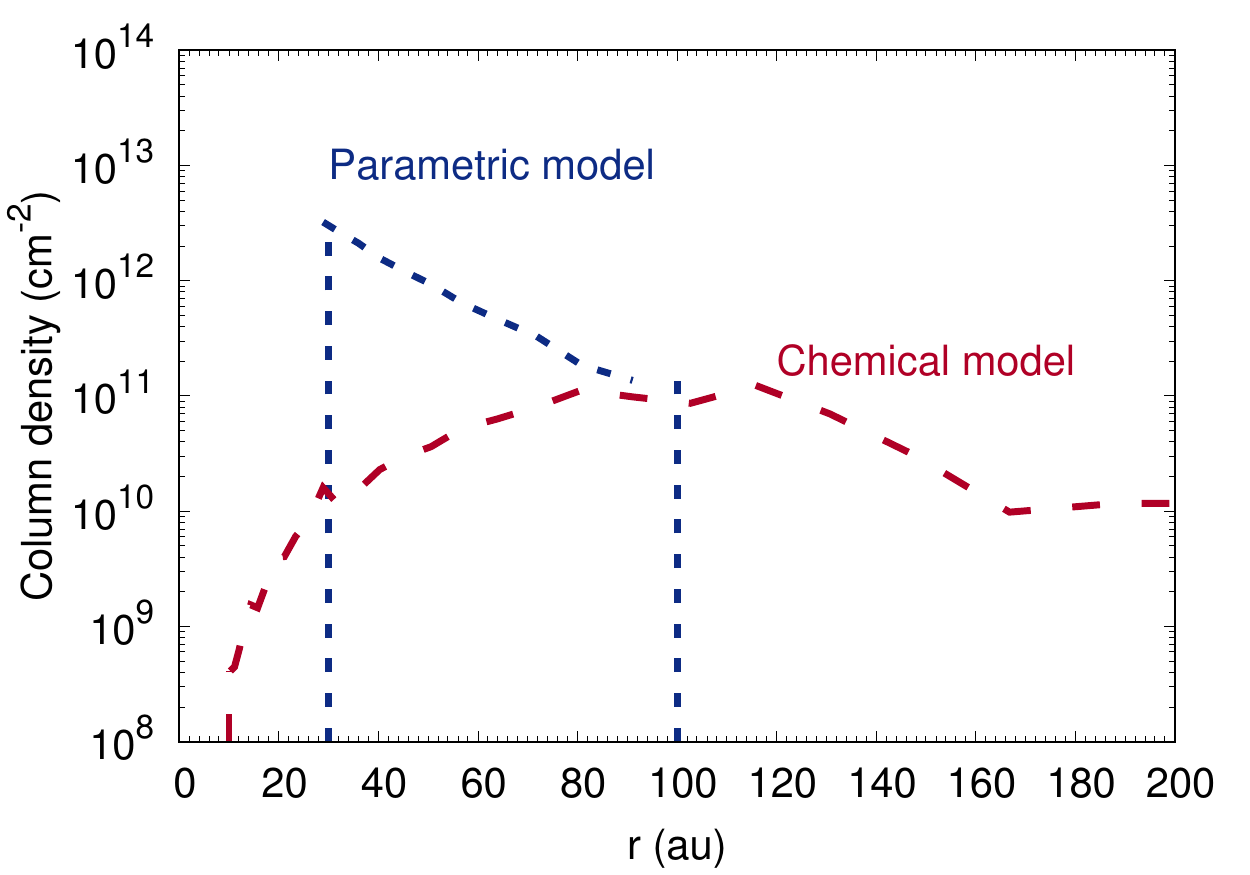}
\end{center}
\caption{Column density of methanol gas as a function of radius for the parametric model 
from \citet{walsh16} and the chemical model (blue and red lines respectively).}
\label{fig4}
\end{figure}

Figure~\ref{fig4} shows the column density of gas-phase methanol as a function of radius 
from the full chemical model compared with the column density derived from the parametric 
models used in \citet{walsh16}.  
The peak column density for the chemical model is $\sim 10^{11}$~cm$^{-2}$ and is reached 
farther out in the disk ($\approx 80-120$~au) than that derived from the parametric models.  
Given that the line emission is optically thin and in local thermodynamic equilibrium 
(LTE), this suggests that the emission peak for the 
chemical model also resides at this location which is a factor of $\approx 3$ 
farther from the star than suggested by the observations. 
\citet{parfenov17} find similar results.  

\section{On the distribution of methanol gas in TW Hya}

The results for TW Hya using the full and updated chemical model show that 
the peak abundance of gas-phase methanol matches well that constrained by the observations. 
Thus, the inclusion of fragmentation upon photodesorption of methanol ice 
helps to reconcile the relatively high abundances of gas-phase methanol 
predicted by the previous generation of protoplanetary disk 
models with complex chemistry 
\citep[$\sim 10^{-10} - 10^{-9}$ with respect to \ce{H2}, e.g.,][]{semenov11,furuya14,walsh14}.  
However, the modelled radial distribution does not agree with the observations: 
the peak emission is predicted at a radius that is 
a factor of $2-3$ beyond that seen in the ALMA data \citep[see also][]{parfenov17}.  
This suggests that the models, despite their complexity, are missing sources of gas-phase 
methanol in the inner regions of protoplanetary disks beyond those already considered.  
\citet{salinas16} speculate that collisions between icy bodies in the 
disk midplane may explain the observed abundance ratio of gas-phase \ce{NH3} 
to gas-phase \ce{H2O} observed in TW~Hya with \emph{Herschel}.  
It remains to be tested whether such a mechanism would yield a 
detectable amount of gas-phase methanol and also if larger molecules like 
methanol would survive such a process.  
\citet{oberg17} analysed ALMA observations of formaldehyde (\ce{H2CO}) in TW Hya 
and found two reservoirs: an inner warm component within $\approx 10$~au and 
an outer colder component beyond $\approx 15$~au arising from the disk atmosphere.  
Given the strong chemical association between methanol and formaldehyde 
it is possible that these two molecules share a similar emission morphology; 
however, methanol is less volatile than formaldehyde. 
Hence, a chemical process that is able to release methanol in this cold region  
is still required.  
One such process is co-desorption with a less volatile yet abundant ice such as CO.   
This has recently been investigated in experiments by \citet{ligterink17}.  
Gas-phase formation of methanol may also need to be revisited.  

An alternative explanation is that the ice reservoir may not be 
present throughout the full radial extent of the disk midplane.  
Clues pointing to this are present in recent high spatial resolution ALMA images of dust 
continuum emission from TW Hya which show that the $\sim$~mm-sized dust grains 
are confined to $\lesssim 60$~au \citep{andrews16,nomura16}.  
This dust morphology shows that the large dust grains in TW Hya 
have settled to the midplane and have drifted inwards due to the head wind 
and resultant loss in angular momentum induced by the velocity differences in gas and mm-sized 
dust in the outer disk \citep[][]{weidenschilling77}.  
Complementary observations of line emission from both single-dish observations and 
spatially resolved observations with ALMA have provided additional support that these 
large dust grains are host to the bulk of the ice reservoir 
(see the recent studies by \citealt{kama16}, \citealt{salinas16}, 
\citealt{bergin16}, and \citealt{du17}).  
Sequestration and concentration of the ice reservoir towards the inner disk midplane 
($\lesssim 60$~au) may also help to explain the emission morphology of 
gas-phase methanol as seen with ALMA.  
This process would help increase the volume of methanol ice in the inner midplane 
($\lesssim 60$~au) and would 
yield an emission morphology which peaks closer to the central star.

\begin{figure}[!h]
\begin{center}
\includegraphics[width=\textwidth]{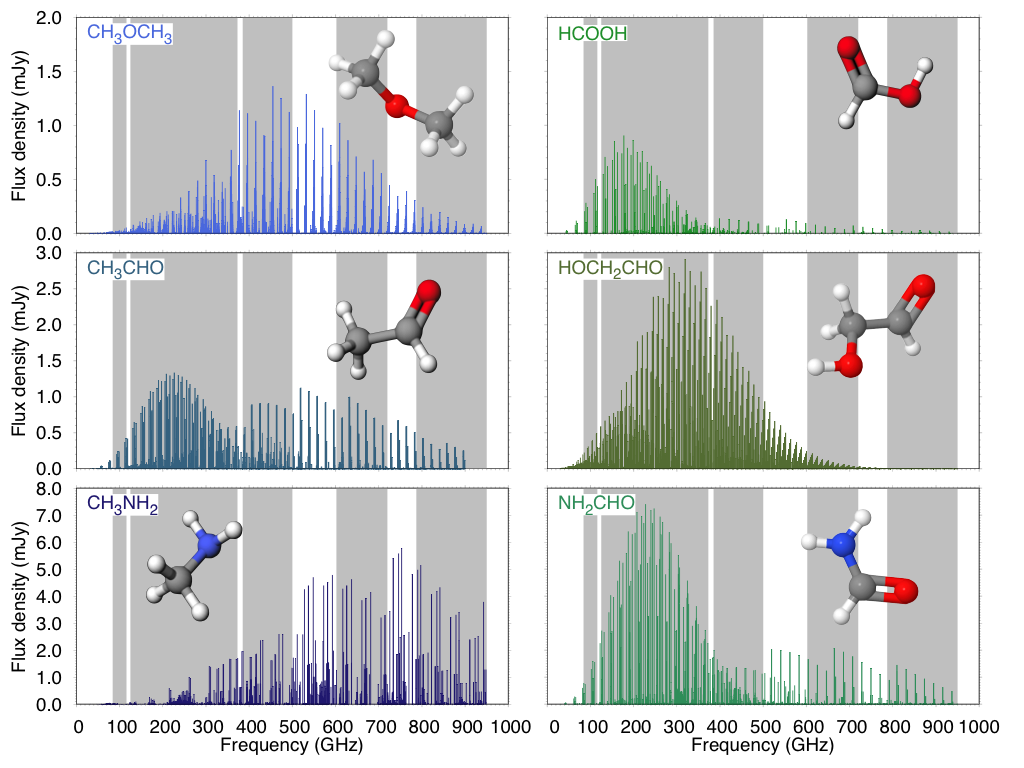}
\end{center}
\caption{Disk-integrated rotational spectra of six COMs generated using the physical model 
of TW Hya and confining the gas-phase abundance to the midplane $z/r \lesssim 0.1$ and 
within $30 \le r \le 100$~au. The assumed abundance 
is that which best fits the TW Hya methanol data, $3\times10^{-12}$ 
with respect to \ce{H2}. The gray shaded regions are the current ALMA observing bands. 
Line lists are from the Cologne Database for Molecular Spectrocopy 
(\texttt{http://www.astro.uni-koeln.de/cdms/}) or JPL (\texttt{https://spec.jpl.nasa.gov/}).}
\label{fig5}
\end{figure}

\section{Detection of COMs higher up the ladder of complexity}

The possibility to detect COMs of a similar or higher complexity as methanol in disks 
around young low-mass stars like TW Hya can now 
be better quantified in light of the detection of methanol in this source. 
The disk-integrated rotational spectra of multiple COMs across the full ALMA 
frequency range have been computed assuming LTE 
holds, that the disk is face on to the line of sight 
\citep[as also done in][]{walsh14}, and that the disk is at the distance 
of TW Hya (54~pc).  
The assumption of LTE for methanol emission was recently tested and 
verified for TW Hya in \citet{parfenov16,parfenov17}.  
The parametric model which reproduces the methanol detection 
in TW Hya is used as a template \citep{walsh16}.  
The methanol is confined to the  
midplane ($z/r \lesssim 0.1$) between a radial range 
of $30-100$~au: this resulted in a best-fit methanol abundance of $3 \times 10^{-12}$ 
with respect to \ce{H2}.  
It is assumed that each COM has this abundance in 
the computation of the spectra.  
This is an optimistic estimate because in other astrophysical environments  
gas-phase methanol is often the most abundant COM.  
This also assumes that all COMs have a similar production mechanism 
to methanol in the disk.

Figure~\ref{fig5} shows the disk-integrated rotational spectra for six COMs:
dimethyl ether (\ce{CH3OCH3)}, formic acid (HCOOH), acetaldehyde (\ce{CH3CHO}), 
glycolaldehyde \\ 
(\ce{HCOCH2CHO}), methylamine (\ce{CH3NH2}), and formamide (\ce{NH2CHO}).  
The spectra are approximately ordered by peak line flux from top to bottom.  
The grey shaded regions show the ALMA observing bands available in ALMA
Cycle 5 (the current observing cycle). 
The frequency of the strongest rotational transitions vary according the 
properties of the molecule \citep[see, e.g.,][]{herbst09}. 
Formic acid, acetaldehyde, and formamide 
all have their strongest transitions towards the lower frequency bands (4, 5, and 6), 
glycolaldehyde peaks in band 7, and dimethyl ether and 
methylamine have their strongest transitions in band 8 and bands 9 and 10, respectively.  
Because of the increase in ALMA sensitivity towards the lower frequencies (for a fixed  
integration time), formamide is a strong contender for detection in protoplanetary disks 
with a peak line flux density of $\approx 7$~mJy.  
Recall that the rms noise of the ALMA data for TW Hya was 
2~mJy beam$^{-1}$ channel$^{-1}$.  
Detections of individual transitions of formic acid, acetaldehyde, 
and glycolaldehyde may also be possible with a sufficiently deep integration.  
However, dimethyl ether and methylamine may lie beyond the capabilities of the 
telescope, at least in candidate disks around young low-mass stars.

These line spectra are computed for a disk at the same distance as 
TW Hya (54~pc). For disks at a distance of $\approx 150$~pc which is more typical 
of nearby star-forming regions, the line emission will be diluted by a factor 
of $\approx 9$.  
However, each of the species considered here has a very high density of 
spectral lines within the observing band within which the peak emission 
is attained.  
For example, formamide has 660 transitions (with $\log(I_\nu)\gtrsim -6$) lying 
between the lower edge of band 6 (211~GHz) and the upper edge of band 7 (373 GHz).  
Hence, stacking of line data from a single species using data from 
unbiased spectral line surveys may enable a signal-to-noise boost of $\sqrt{n}$ 
where $n$ is the number of stacked lines.  
It should be noted that care must be taken to avoid stacking transitions of vastly different
excitation energies and checks for potential line blends will be required.  
On this latter point, the typical line widths of disk-integrated line profiles 
from protoplanetary disks around low-mass stars is of the order of a few km~s$^{-1}$; 
hence, blending of neighbouring line transitions is expected to be less of an 
issue than it is towards sources in which the observed spectra 
are found to be at or close to the confusion limit (e.g., hot cores, see \citealt{herbst09}).

\end{document}